\documentclass{amsart}
\usepackage{amsfonts}
\usepackage{amssymb}

\newtheorem{theorem}{Theorem}[section]
          
\newtheorem{lemma}[theorem]{Lemma}              
              
\newtheorem{proposition}[theorem]{Proposition}
\newtheorem{corollary}[theorem]{Corollary}

\renewenvironment{proof}{\noindent {\bf Proof: }}{\QED\medskip}
\def\QED{{\hspace*{\fill}{\vrule height 1ex width 1ex }\quad} 
    \vskip 0pt plus20pt}

\newcommand{\Ir}{\mathbb{Z}}
\newcommand{\Cx}{\mathbb{C}}
\newcommand{\Bdd}{\mathcal{B}}
\newcommand{\ket}[1]{\vert{#1}\rangle}
\newcommand{\GS}{\mathcal{G}}
\newcommand{\ip}[2]{\langle{#1}|{#2}\rangle}
\newcommand{\Hil}{\mathcal{H}}
\newcommand{\Proj}{\operatorname{Proj}}
\newcommand{\Obs}{\mathcal{A}}
\newcommand{\sech}{\operatorname{sech}}
\newcommand{\Ker}{\operatorname{Ker}}

\begin{document}

{\huge
\renewcommand{\thefootnote}{\fnsymbol{footnote}}
\noindent
Improved Bounds on the Spectral Gap above 
Frustra-\\
tion Free Ground States of Quantum Spin Chains
\footnote{\noindent Copyright
\copyright\ 2002 by the authors. Reproduction of this article in its entirety, 
by any means, is permitted for non-commercial purposes.} \\[15pt]}
\renewcommand{\thefootnote}{\arabic{footnote}}
\setcounter{footnote}{1}
{\large \bf WOLFGANG L SPITZER$^1$
and SHANNON~STARR$^2$}\\
{$^1$ \it Department of Mathematics,
University of California, Davis,
Davis, CA 95616-8633, USA. email
spitzer@math.ucdavis.edu}\\
{$^2$ \it Department of Physics,
Princeton University,
Princeton, NJ 08544, USA. email
sstarr@math.princeton.edu}\\[7pt]
(9 December 2002)\\[7pt]
\noindent
{\bf Abstract}
In \cite{Nac2,Nac1} Nachtergaele obtained explicit lower bounds
for the spectral gap above many frustration free quantum spin chains
by using the ``martingale method''.
We present simple improvements to his main bounds
which allow one to obtain a sharp lower bound for the
spectral gap above the spin-1/2 ferromagnetic XXZ chain.
As an illustration of the method, we also calculate
a lower bound for the spectral gap of the AKLT model, which
is about 1/3 the size of the expected gap.\\[7pt]

\section{Introduction}
In this paper, we obtain lower bounds on the spectral gap above
frustration free ground states, which are special ground states
which exist for certain special quantum spin systems.
In the simplest case, a ``frustration free'' quantum spin chain
can be defined as follows:
For each site, let $\Hil_x \equiv \Cx^d$ be a finite dimensional
Hilbert space, such that all $\Hil_x$ are equivalent to one another.
For any finite subset $\Lambda \subset \Ir$, one defines
$\Hil_\Lambda = \bigotimes_{x \in \Lambda} \Hil_x$.
Consider an interaction
$\Phi$, defined on all finite subsets of $\Ir$,
such that $\Phi(\Lambda_0) \in \Bdd(\Hil_{\Lambda_0})$
for all finite $\Lambda_0$.
Then for any finite subset $\Lambda \subset \Ir$, we define
$$
H(\Lambda) = \sum_{\Lambda_0 \subset \Lambda} \Phi(\Lambda_0)\, ,
$$
where, as usual, $\Phi(\Lambda_0) \in \Bdd(\Hil_{\Lambda_0})$ is extended
to $\Hil_{\Lambda}$ by tensoring with the identity on 
$\Hil_{\Lambda \setminus \Lambda_0}$.
This is the set-up for a typical quantum spin system.
What distinguishes a frustration free quantum spin system are two assumptions
\begin{enumerate}
\item For each $\Lambda_0$, $\Phi(\Lambda_0)$ is a positive operator.
\item For each nonempty $\Lambda$, the subspace $\GS(\Lambda) = \Ker(H(\Lambda))$
satisfies $\dim(\GS(\Lambda)) > 0$.
\end{enumerate}
In this case, we call $\GS(\Lambda)$ the ground state space, and declare the ground
states to be frustration free.

We point out an obvious implication.
Since all $\Phi(\Lambda_0)$ are positive,
$$
\Ker(H(\Lambda)) = \cap_{\Lambda_0 \subset \Lambda} \Ker(\Phi(\Lambda_0))\, .
$$
This means that any ground state $\psi \in \GS(\Lambda)$ not only minimizes
the total energy, i.e.\ is a minimum energy eigenvector of $H(\Lambda)$,
it also minimizes all the local interaction energies.
This is to be contrasted with a frustrated spin system.

To mention just a few interesting examples of frustration free systems,
there are ferromagnets: the isotropic Heisenberg ferromagnet has a frustration
free ground state, as do the anisotropic ``kink'' ferromagnets known as XXZ models.
A frustration free antiferromagnet is the AKLT model, invented by Affleck, Kennedy,
Lieb and Tasaki \cite{AKLT1}, which is an important model
for understanding the role of spin-dimension in antiferromagnetic chains.
(Haldane~\cite{Haldane1} conjectured that the antiferromagnetic, isotropic 
Heisenberg chain has gapless excitations above the ground state if the spin is half-integer 
but is gapfull if the spin takes an integer value.)
There are numerous stochastic models, such that the Markov generators can be viewed
as frustration free models.
Of these, we just mention one, the Kac model \cite{Kac}, for which the gap has recently 
been computed exactly by Carlen, 
Carvalho and Loss \cite{CCL} using a technique, similar in spirit, but different than the martingale
method, which is specific to mean field models.
We point this out to indicate that this is still a very active field of research.

Henceforth we will assume that $\Phi$ is only a nearest neighbor interaction, and 
that it is translation invariant.
Neither of these requirements is essential.\footnote{An example of a non-translation
invariant frustration free model was treated in \cite{CNS}. There, the XXZ model is perturbed by
a defect, which nevertheless is still frustration free. In that paper, the ``martingale method''
is applied in a slightly different way because there the spin chain is grown on both edges,
while we will grow the spin chain along one edge only.}
This allows us to write
$H_{\Lambda} = \sum_{x,x+1 \in \Lambda} h(x,x+1)$, where $h(x,x+1)$ is a fixed positive operator
of just two spins.

\section{Main Theorem}
The main result of this paper is a method of calculating lower
bounds for the spectral gap, by diagonalizing a family of finite dimensional
matrices, with dimensions much smaller than the starting
Hamiltonians $H_{\Lambda}$, $\Lambda \subset \Ir$.
There are two situations where this is particularly useful: First,
if there is a symmetry which allows explicit diagonalization of these
matrices, which is the case for the spin-$1/2$ XXZ model.
Second, for the finitely correlated states (see Section \ref{Eg2AKLT})
the matrices involved
are all of a fixed finite dimension.
If the dimension is small enough, the matrices can be diagonalized
by hand, which is the case for the AKLT model.

Suppose $\Lambda = [a,b]$ and $\Lambda_0 = [c,d] \subset \Lambda$.
Then we define $H(c,d)=H_{\Lambda_0}$, interpreted as an operator on
$\Hil_\Lambda$.
We suppress the dependence on $\Lambda$ from the notation.
Similarly we define $\GS(c,d) = \Ker H(c,d)$.
We define $G(c,d) = \Proj (\GS(c,d))$ to be the orthogonal projection
onto $\GS(c,d)$.
The spectral gap, $\gamma(c,d)$, is defined to be the largest number such that
$$
H(c,d) \geq \gamma(c,d) (1 - G(c,d))\, .
$$
This number is independent of $\Lambda$, and depends only on $c-d+1 = \#[c,d]$.
We define $\gamma_N = \min_{2\leq n\leq N} \gamma(1,n)$,
which has the useful property of being a nonincreasing function of $N$.
A sensible convention is that
$H(a,a)=0$, $\GS(a,a) = \Hil_{a} = \Cx^d$, and $G(a,a) = 1$. 
However, there is no way to define $\gamma(a,a)$,
since both $H(a,a)$ and $1-G(a,a)$ vanish.

The auxiliary Hamiltonian, whose analysis
is easier than that of $H_\Lambda$ and which gives
information on $\gamma(a,b)$, is just the ground state projector
restricted to an appropriate subspace.
Suppose $\Lambda_0 = [-m,n] \subset \Lambda = [a,b]$.
Define
\begin{equation}
\label{varepsemen}
\varepsilon(m,n) = \sup  \{\ip{\psi}{G(0,n) \psi} 
	: \psi \in \mathcal{G}(-m,0)\, ,\
	\psi \in \mathcal{G}(-m,n)^\perp\, ,\ \|\psi\| = 1\}\, .
\end{equation}
The value of $\varepsilon(m,n)$ is independent
of $a$ and $b$ as long as $a\leq -m$ and $b\geq n$, and by translation
symmetry the answer would be the same if the interval $[-m,n]$ were
replaced by $[x-m,x+n]$ for any $x$.
For $m,n \geq 1$, we define
$$
\varepsilon_{m,n}  =  \sup_{m' \geq m} \varepsilon(m',n)\, ,
$$
which is monotone decreasing in $m$, but not necessarily in $n$.

Our main result is the following:

\begin{theorem}
\label{MM}
\label{MnThm}
Suppose that $\varepsilon_{m,n} < 1/2$, for some $n \geq m\geq 1$.
For any $N \geq 2$
and $\psi \in \GS(1,N)^\perp$, define
$$
\psi_1 = (1-G(1,N-n)) \psi\, ,\ 
\psi_2 = G(1,N-n) \psi\, ,
$$
whence it follows $\psi = \psi_1 + \psi_2$.
(If $N\leq n+1$ then $\psi_1=0$, $\psi_2=\psi$.)
Then 
\begin{equation}
\label{ThmEqn}
\ip{\psi}{H(1,N) \psi} 
  \geq \gamma_{m+n} (\alpha(\varepsilon_{m,n}) \|\psi_1\|^2
  + \beta(\varepsilon_{m,n}) \|\psi_2\|^2)\, ,
\end{equation}
where 
$\alpha(\varepsilon) = (\sqrt{1-\varepsilon}-\sqrt{\varepsilon})^2$ and
$\beta(\varepsilon) = \sqrt{1 - \varepsilon}(\sqrt{1-\varepsilon} - \sqrt{\varepsilon})$.
\end{theorem}

Our improved bound is due to the following simple estimate.

\begin{lemma}
\textit{(a)} Suppose $H$ is a positive matrix,
$G = \Proj(\Ker(H))$, and $\gamma$ is the gap; so that 
$H \geq \gamma (1-G)$.
If $\phi$ is a vector which is not in the kernel of $H$ --
not necessarily orthogonal to the kernel either --
then for any $\psi$
\begin{equation}
\label{Vartnl}
\ip{\psi}{H \psi} \geq \gamma |\ip{\psi}{(1-G) \phi}|^2/\ip{\phi}{(1-G)\phi}\, .
\end{equation}
\textit{(b)} If $G$ is an orthogonal projection, $\phi$ is a nonzero vector, and $\psi$
is any vector orthogonal to $\phi$, then
\begin{equation}
\label{Vartn2}
|\ip{\psi}{G \phi}|^2 \leq \|\psi\|^2 \|\phi\|^2 \langle{G}\rangle_\phi (1 - \langle{G}\rangle_\phi)\, ,
\end{equation}
where
$\langle{G}\rangle_\phi = \ip{\phi}{G \phi}/\|\phi\|^2$.
\end{lemma}

\begin{proof}
Both are by Cauchy-Schwarz.
First,
\begin{align*}
|\ip{\psi}{(1-G) \phi}|^2 
  = |\ip{(1-G)\psi}{(1-G) \phi}|^2
  &\leq \|(1-G)\psi\|^2 \|(1-G)\phi\|^2 \\
  &\leq \gamma^{-1} \ip{\psi}{H \psi} \|(1-G) \phi\|^2\, .
\end{align*}
For the second, choose an orthonormal basis $\phi_1,\phi_2,\dots$ with
$\phi_1 = \phi/\|\phi\|$. Then
\begin{align*}
|\ip{\psi}{G \phi}|
  \leq \sum_{n\geq 2} |\ip{\psi}{\phi_n} \ip{\phi_n}{G \phi}| 
  &\leq \Big(\sum_{n\geq 2} |\ip{\psi}{\phi_n}|^2\Big)^{1/2}
  \Big(\sum_{n\geq 2} \ip{G \phi}{\phi_n} \ip{\phi_n}{G\phi}\Big)^{1/2} \\
  &= \|\psi\| \Big(\sum_{n\geq 2} \ip{G \phi}{\phi_n} \ip{\phi_n}{G\phi}\Big)^{1/2}\, .
\end{align*}
Moreover,
$$
\sum_{n \geq 2} \ip{G \phi}{\phi_n} \ip{\phi_n}{G \phi}
  = \ip{G \phi}{G\phi} - \ip{G \phi}{\phi_1} \ip{\phi_1}{G\phi}
  = \|G \phi\|^2(1 - \|G\phi\|^2/\|\phi\|^2)\, .
$$
\end{proof}

\begin{proof}(Of Theorem \ref{MnThm})
The proof is by induction.
The first step is to check that the proposition is trivially
satisfied for $N \leq m+n$.
Indeed, since $\alpha \leq \beta \leq 1$, we have
$$
\gamma(N) \geq \gamma_{m+n} \geq \max(\alpha \gamma_{m+n},\beta \gamma_{m+n})\, ,
$$
by the definition of $\gamma_{m+n}$.
The next step is the induction step.

For the induction hypothesis we suppose the proposition is true whenever 
$N\leq N_0$, where $N_0 \geq m+n$, and then prove the proposition is
also true for $N=N_0+1$.
This means we must show
\begin{equation}
\label{First Desideratum}
\ip{\psi}{H(1,N) \psi} \geq \gamma_{m+n} [\alpha \|\psi_1\|^2 + \beta \|\psi_2\|^2]
\end{equation}
where $G(1,N) \psi =0$,
$\psi_1 = [1-G(1,N-n)] \psi$ and
$\psi_2 = [G(1,N-n) - G(1,N)] \psi$. 
It is useful to make a further orthogonal decomposition  
$$
\psi_{11} = [1-G(1,N-2n)] \psi_1\, ,\ 
\psi_{12} = [G(1,N-2n) - G(1,N-n)] \psi_1\, .
$$
(We have assumed that $N>2n$, otherwise we set $\psi_{11}=0$.)
By the induction hypothesis, 
we conclude
$$
\ip{\psi_1}{H(1,N-n) \psi_1}
  \geq \gamma_{m+n} [\alpha \|\psi_{11}\|^2 + \beta \|\psi_{12}\|^2]\,.
$$
Since $\psi_2 \in \GS(1,N-n)$, it is clear that 
$\ip{\psi}{H(1,N-n) \psi} = \ip{\psi_1}{H(1,N-n) \psi_1}$.
So to prove equation (\ref{First Desideratum}), we just need to prove
\begin{equation}
\label{Second Desideratum}
\ip{\psi}{H(N-n,N) \psi} \geq \gamma_{m+n} 
  (\beta \|\psi_2\|^2 - (\beta - \alpha) \|\psi_{12}\|^2)\, .
\end{equation}
We assume
\begin{equation}
\label{Ass1}
\|\psi_2\|^2 
  > \Big(1 - \frac{\alpha}{\beta}\Big) \|\psi_{12}\|^2
  = \frac{\sqrt{\varepsilon}}{\sqrt{1-\varepsilon}} \|\psi_{12}\|^2
\end{equation}
because otherwise equation (\ref{Second Desideratum}) is automatically satisfied.
(Henceforth, we will write $\varepsilon$ in place $\varepsilon_{m,n}$.)

To estimate the left hand side of equation (\ref{Second Desideratum}),
we use Lemma 2.4.
Specifically, by part (a), substituting $\psi_2$ for $\phi$, we have
$$
\ip{\psi}{H(N-n,N) \psi} 
  \geq \gamma_{m+n} \frac{|\ip{\psi}{[1-G(N-n,N)] \psi_2}|^2}{\ip{\psi_2}{[1-G(N-n,N)]\psi_2}}\, ,
$$
where we have used $\gamma_{m+n} \leq \gamma_{n+1}$.
We define the number
$$
\eta = \langle{G(N-n,N)}\rangle_{\psi_2} = \ip{\psi_2}{G(N-n,N) \psi_2}/\|\psi_2\|^2
$$
which allows us to rewrite 
\begin{equation}
\label{First Bound}
\ip{\psi}{H(N-n,N) \psi} 
  \geq \gamma_{m+n} |\ip{\psi}{[1-G(N-n,N)] \psi_2}|^2\Big/(1-\eta) \|\psi_2\|^2\, .
\end{equation}
The denominator is not zero because, by (\ref{Ass1}), $\psi_2 \neq 0$,
and by the definition of $\varepsilon_{m,n}$,
$\eta \leq \varepsilon_{m,n} < 1/2$. 

Writing $G$ in place of $G(N-n,N)$,
\begin{equation}
\label{First Little Equation}
\ip{\psi}{[1-G] \psi_2}
  = (1-\eta) \|\psi_2\|^2 - \ip{\psi_{11}}{G \psi_2} - \ip{\psi_{12}}{G \psi_2}\, .
\end{equation}
It is easy to see that
\begin{equation}
\label{Second Little Equation}
\ip{\psi_{11}}{G \psi_2} = \ip{\psi_{11}}{G(N-n,N) \psi_2} = 0
\end{equation}
because $\psi_{11}$ and $G(N-n,N) \psi_2$ are both eigenvectors of the Hermitian
operator $G(1,N-2n)$, with different eigenvalue.
(Note that $G(N-n,N)$ commutes with $G(1,N-2n)$ because they are localized on
disjoint intervals.)
By Lemma 2.4 (b),
\begin{equation}
\label{Third Little Equation}
\begin{split}
|\ip{\psi_{12}}{G \psi_2}|
  &\leq \|\psi_{12}\| \cdot \|\psi_2\| \sqrt{\eta(1-\eta)}\, .
\end{split}
\end{equation}
Combining equations (\ref{First Little Equation}), (\ref{Second Little Equation})
and (\ref{Third Little Equation}), gives
\begin{equation}
\label{fourth}
\ip{\psi}{[1-G] \psi_2} 
  \geq \sqrt{1-\eta} \|\psi_2\| ( \sqrt{1-\eta} \|\psi_2\| - \sqrt{\eta} \|\psi_{12}\|)\, .
\end{equation}
Using (\ref{Ass1}) and observing that $\frac{\eta}{1-\eta} \leq \frac{\sqrt{\varepsilon}}{\sqrt{1-\varepsilon}}$,
the right hand side of (\ref{fourth}) is positive.
Hence
\begin{equation}
\label{Little Conclusion}
|\ip{\psi}{[1-G]\psi_2}|^2 \geq (1-\eta) \|\psi_2\|^2 
  ( \sqrt{1-\eta} \|\psi_2\| - \sqrt{\eta} \|\psi_{12}\|)^2\, .
\end{equation}
Therefore, by (\ref{First Bound}), equation (\ref{Second Desideratum})
would follow from
\begin{equation}
( \sqrt{1-\eta} \|\psi_2\| - \sqrt{\eta} \|\psi_{12}\|)^2
  \geq \beta \|\psi_2\|^2 - (\beta - \alpha) \|\psi_{12}\|^2\, .
\end{equation}

The proof is reduced to checking that the bilinear form
$$
B(x,y) = (\beta - \alpha + \eta) x^2 - 2 \sqrt{\eta(1-\eta)} x y + (1-\eta-\beta) y^2
$$
is positive. The conditions for positivity are
\begin{gather}
\label{Condition A}
\beta - \alpha + \eta \geq 0\, ,\quad
1-\eta-\beta \geq 0\, ,\\
\label{Condition B}
\eta(1-\eta) \leq (1-\eta-\beta)(\beta-\alpha+\eta)\, .
\end{gather}
They must be satisfied for all $\eta$ in the range $[0,\varepsilon]$
(where we write $\varepsilon$ for $\varepsilon_{m,n}$).
Condition (\ref{Condition B}) is equivalent to
$(1-\beta)(\beta - \alpha) \geq  \eta (2\beta-\alpha)$.
By condition (\ref{Condition A}), this is the same as
$$
\eta \leq (1-\beta)(\beta-\alpha)/(2\beta - \alpha)
$$
for all $\eta\leq \varepsilon$.
However the inequality with $\eta = \varepsilon$ subsumes all
those with $\eta<\varepsilon$. 
Again using (\ref{Condition A}), this is equivalent to
$$
\alpha \leq \beta(1-2\varepsilon-\beta)/(1-\varepsilon-\beta)\, .
$$
The maximum value of $\alpha$ satisfying these constraints occurs when
$$
\alpha = (\sqrt{1-\varepsilon} - \sqrt{\varepsilon})^2\, ,\quad
\beta = 1 - \varepsilon - \sqrt{\varepsilon(1-\varepsilon)}\, ,
$$
which is just what we set out to prove.
\end{proof}

The previous theorem can be used to obtain uniform bounds
for the spectral gap above all finite volume ground states.
The situation for infinite volume ground states is usually more
complicated than just taking the limit of the finite volume gaps.
An infinite volume ground state is a state $\omega$ on the quasilocal
observable algebra $\Obs_0$ which is locally stable in the
sense that for any finite volume $\Lambda$, and any
$X \in \Obs_\Lambda$, $\omega(X^* \delta(X)) \geq 0$,
where $\delta$ is the derivation 
$\delta(X) = \lim_{\Lambda'\nearrow\Ir} [H_{\Lambda'},X]$.
For such a ground state, one can define the gap $\gamma(\omega)$ to be
the greatest number $\gamma$ satisfying
$$
\omega(\delta(X)^* \delta^2(X)) \geq \gamma \,\omega(\delta(X)^* \delta(X))\, ,
$$
for any local observable $X \in \Obs_\Lambda$, with $\Lambda$ finite.
The purpose of taking three powers of $\delta$ on the left hand side is to restrict
to the positive energy states, since many ground states may be quasilocal
perturbations of one another.
With this definition, we can state the main corollary. 

\begin{corollary}
\label{MnCor}
Suppose that $\varepsilon_{m,n} < 1/2$, for some $n \geq m\geq 1$.
Then there is a uniform lower bound on the finite volume spectral gaps
$$
\inf_{N\geq 2} \gamma(N) \geq \gamma_{m+n} (1 - 2 \sqrt{\varepsilon_{m,n}(1-\varepsilon_{m,n})})\, .
$$
Moreover, if $\omega$ is an infinite volume ground state, obtained
as a weak-$*$ limit of finite volume ground states, then $\gamma(\omega)$
obeys the same bound.
\end{corollary}

\begin{proof}
The proof of the first part follows trivially from
Theorem \ref{MnThm}.
One simply observes that $\beta(\varepsilon)>\alpha(\varepsilon)$ and 
$\alpha(\varepsilon)=1-2\sqrt{\varepsilon(1-\varepsilon)}$.

The proof of the second part follows an argument of Koma and Nachtergaele
\cite{KN1}, which was stated for the XXZ model.
Suppose $\omega$ is a weak-$*$ limit of finite volume ground states
$\omega_N(\cdots) = \ip{\psi_N}{\cdots \psi_N}$, where the unit vectors
$\psi_N$ are each in $\Hil_{\Lambda_N}$ for $\Lambda_N \nearrow \Ir$.
Then for any local $X\in\Obs_\Lambda$, there exists an $N_0$ with $X \in \Obs_{\Lambda_N}$
for all $N\geq N_0$, and
$\omega(X) = \lim_{N \to \infty} \omega_N(X)$, the limit starting at $N=N_0$.
Now note that $\delta(X)^* \delta^2(X)$ is also local, localized in
$\Lambda^{(2)}$ consisting of all points no more than 2 units from $\Lambda$.
There is also an $N_0$ such that $\Lambda_N \supset \Lambda^{(2)}$
for all $N>N_0$, and then we evaluate
$$
\omega_N(\delta(X)^* \delta^2(X)) 
  = \ip{\delta(X) \psi_N}{H(\Lambda_N) \delta(X) \psi_N}\, .
$$
Let $\psi_N' = \delta(X) \psi_N$.
Then if $\phi$ is any ground state of $H=H(\Lambda_N)$, 
$$
\ip{\phi}{\psi_N'} 
  = \ip{\phi}{(HX-XH) \psi_N} = 0 
$$
because $H \psi_N = H \phi = 0$.
So $\psi_N'$ is orthogonal to the ground state space, hence
$$
\omega_N(\delta(X)^* \delta^2(X)) \geq \gamma(N) \,\ip{\psi_N'}{\psi_N'} = 
\gamma(N) \,\omega_N(\delta(X)^* \delta(X))\, .
$$
Using the uniform lower bound for the finite volume gap, and taking liminf's
proves the claim.
\end{proof}

Theorem \ref{MnThm} apparently is a stronger result than the last corollary.
For example, one can use it to prove that, whenever the lower bound on the infinite volume gap 
is achieved, then the first excitation is not an ``edge state''.
The only example we know where the lower bound is attained
is the XXZ model, and there the first excitation is actually not an eigenvalue but
the bottom of a continuous band of spectrum with quasi-momentum 0.
This is an important difference between the XXZ model and the AKLT model,
where the lowest variational excitations (see \cite{FS})
have quasi-momentum equal to $\pi$.

\section{Example 1 : The XXZ model}

The XXZ ferromagnet, with \textit{kink}-inducing boundary fields
is a translation invariant, nearest-neighbor model
with anisotropic exchange 
$$
h(n,n+1) = -\sech(\xi) \boldsymbol{S}_n \cdot \boldsymbol{S}_{n+1}
	- (1 - \sech(\xi)) S_n^{3} S_{n+1}^{3}
	+ j \tanh(\xi) (S_n^{3} - S_{n+1}^{3})\, .
$$
where $\xi \geq 0$ determines the amount of anisotropy.
The isotropic Heisenberg model is $\xi = 0$, while
$\xi=\infty$ is the Ising model with kink boundary conditions.
For $j=1/2$, the Hamilton is invariant under the action of
the quantum group $\textrm{SU}_q(2)$, where $q = e^{-\xi}$.
In this section we will prove the following proposition.
\begin{proposition}
\label{XXZProp}
Let $\xi>0$. Then the spectral gap above any infinite volume,
weak-$*$ limit of finite volume kink ground states is 
$\gamma(\xi) = 1 - \sech\xi$.
\end{proposition}
The representation of $\textrm{SU}_q(2)$ is generated by three spin 
operators
$$
S^3(1,N) = \sum_{n=1}^N S_n^{3}\, ,\quad
S^+(1,N) = \sum_{n=1}^N e^{-(n-1) \xi} S_n^+\, ,\quad
S^-(1,N) = \sum_{n=1}^N e^{-(N-n) \xi} S_{n}^-\, .
$$
The first is the usual generator of $\textrm{U}(1)$; the other
two are deformations of the spin raising and lowering operators for $\textrm{SU}(2)$.
Notably, they are not permutation invariant.
For us, the fundamental feature is a decomposition property.
Define $S^{3,+,-}(a,b) = \tau^{a-1}(S^{3,+,-}(1,b+1-a))$,
where $\tau$ is the translation automorphism.
Then for any $1\leq n\leq N$
\begin{align*}
S^+(1,N) &= S^+(1,n) + e^{-n\xi} S^+(n+1,N)\, ,\\
S^-(1,N) &= e^{-(N-n)\xi} S^-(1,n) + S^-(n+1,N)\, .
\end{align*}

The representation theory $\textrm{SU}_q(2)$ for $q>0$ is identical to
that of $\textrm{SU}(2)$ in many respects.
In particular, the Clebsch-Gordon series is the same.
This means the irreducible representations of each dimension, or multiplets,
occur with the usual multiplicities in $\Hil(1,N)$.
Particularly, there is a unique maximal dimensional multiplet.
As in the isotropic case, 
it is known \cite{PS, ASW, GW} that the ground state subspace of the XXZ ferromagnet 
is precisely this maximal dimensional multiplet.
Moreover, it was shown \cite{KN2} that all the infinite volume ground states
are obtained as weak-$*$ limits of these states, which means Proposition \ref{XXZProp}
is particularly relevant.
We refer to a generic $n$-dimensional $\textrm{SU}_q(2)$ multiplet as $V(n)$.

To prove the proposition, we use Corollary \ref{MnCor}.
Nachtergaele already calculated every $\varepsilon(m,1)$ in \cite{Nac2}.
We will repeat his argument, for completeness.
This also serves as a slightly simpler preparation for the calculation in the next
section.
A major motivation for this letter is the fact that Nachtergaele's calculation,
with our improved bound, is sufficient to calculate the exact gap.

We now state a lemma which is generally helpful for calculating $\varepsilon(m,n)$.
\begin{lemma}
\label{UsefulLemma}
The Hermitian operator $K(-m,n)=G(-m,0) G(0,n) G(-m,0)$ has range contained
in the Hilbert space $\GS(-m,0) \otimes \GS(1,n)$.
The eigenspace for eigenvalue $1$ is precisely $\GS(-m,n)$.
The number $\varepsilon(m,n)$ is the largest eigenvalue less than $1$.
\end{lemma}
\begin{proof}
The first statement is true because 
$G(0,n) G(-m,0)$ has range contained in $\GS(1,n)$, and $[G(-m,0),G(1,n)]=0$
implies $K(-m,n)$ also has range contained in $\GS(1,n)$. 
Clearly $\GS(-m,n)$ is an eigenvalue-1 eigenspace of $K(-m,n)$.
If $K(-m,n)\psi=\psi$, then $G(-m,0) \psi = \psi$
and $\ip{\psi}{G(0,n) \psi} = 1$, which implies $G(0,n) \psi = \psi$.
Since $H$ is a nearest-neighbor Hamiltonian, this implies $\psi \in \GS(-m,n)$.
The last statement follows from the minimax principle.
\end{proof}

\begin{proof}(Of Proposition~\ref{XXZProp})
In the case of the XXZ model, the space $\GS(-m,0) \subset \Hil(-m,0)$ is 
the unique $(m+2)$-dimensional multiplet $V(m+2)$.
The space $\GS(1,1)$ is just $V(2)$.
By Clebsch-Gordon, $V(m+2) \otimes V(2) = V(m+3) \oplus V(m+1)$.
Since $G(-m,0)$ and $G(0,1)$ are $\textrm{SU}_q(2)$ invariant, so is $K(-m,1)$.
This implies that $K(-m,1) = \lambda_1 \Proj(V(m+3)) + \lambda_2 \Proj(V(m+1))$.
Of course, since $V(m+3)$ is the maximal dimensional irreducible
multiplet in $\Hil(-m,1)$, we know $\lambda_1=1$, by the lemma.
We can restrict to the two-dimensional (invariant) subspace of $\GS(-m,0)\otimes \GS(1,1)$,
which is the intersection with the eigenspace $\{\psi : S^3(-m,1) \psi = m/2 \psi\}$.
Then it is easy to determine $\lambda_2$ from the trace.
In the following we will use a nonorthogonal basis, which simplifies the 
algebra a bit.

Every vector can be written as
$$
\psi = (\alpha S^-(-m,0) + \beta e^{-(m+1) \xi} S^-(1,1) \ket{\Omega}
$$
for some $\alpha$, $\beta$, where
$\ket{\Omega}$ is the all up-spins state.
Note $S^-(1,1)$ is the spin lowering operator for the interval $[1,1] = \{1\}$;
i.e., it is the spin lowering operator at the site $1$.
We observe that
$$
\psi = (\alpha S^-(-m,-1) + \alpha e^{-m\xi} S^-(0,0) + \beta e^{-(m+1)\xi} S^-(1,1) \ket{\Omega}\, .
$$
Since $S^-(0,1) = S^-(0,0) + e^{-\xi} S^-(1,1)$, it is clear that
$$
G(0,1) \psi = \Big(\alpha S^-(-m,-1) 
	+ \frac{\alpha + \beta e^{-2\xi}}{1+e^{-2\xi}} e^{-m\xi} (S^-(0,0) + e^{-\xi} S^-(1,1))\Big) 
	\ket{\Omega}\, .
$$
Similarly, since $S^-(-m,0) = S^-(-m,-1) + e^{-m\xi} S^-(0,0)$, we have
\begin{align*}
G(-m,0) G(0,1) \psi
   &= \Big((Z+e^{-2m\xi})^{-1}(\alpha Z + e^{-2m\xi} \frac{\alpha + e^{-2\xi}\beta}{1+e^{-2\xi}})
     S^-(-m,0)\\
   &\qquad + \frac{\alpha + e^{-2\xi}\beta}{1+e^{-2\xi}} e^{-(m+1)\xi} S^-(1,1)\Big) \ket{\Omega}
\end{align*}
where $Z = \|S^-(-m,-1) \ket{\Omega}\|^2 = (1-e^{-2m\xi})/(1-e^{-2\xi})$.
(We are exploiting the fact that $G(0,1)$ and $G(-m,0)$ are rank-one projections restricted to the
appropriate subspace.)
Hence the eigenvalues are equal to the eigenvalues of the matrix
$$
\frac{1}{(Z+e^{-2m\xi})(1+e^{-2\xi})}
\begin{bmatrix} 
  Z (1+e^{-2\xi}) + e^{-2m\xi} & e^{-2(m+1)\xi} \\
  (Z+e^{-2m\xi}) & e^{-2\xi} (Z+e^{-2m\xi})
\end{bmatrix}
$$
The trace of this matrix is $1\, +\, e^{-2\xi} Z/((Z+e^{-2m\xi})(1+e^{-2\xi}))$.
Since one of the eigenvalues is 1, that means the other eigenvalue is
$$
\varepsilon(m,1)
  = \frac{e^{-2\xi} Z}{(Z+e^{-2m\xi})(1+e^{-2\xi})}
  = \frac{e^{-2\xi}}{1+e^{-2\xi}} \cdot \frac{1-e^{-2m\xi}}{1-e^{-2(m+1)\xi}}\, .
$$
This is increasing with $m$.
So, $\varepsilon_{1,1} = e^{-2\xi}/(1+e^{-2\xi}) < 1/2$ for $\xi > 0$.
Also, $\gamma_2=1$.
So $\gamma_2 (1 - 2 \sqrt{\varepsilon_{1,1} (1-\varepsilon_{1,1})}\,)
  = 1 - \sech \xi$.

By Corollary \ref{MnCor} this gives the lower bound for the gap of Proposition
\ref{XXZProp}.
But it is also easy to obtain this as an upper bound for the gap above the all-up-spins
state, using a sequence of variational wave function
approximating a long spin-wave state of quasimomentum zero.
Since the all-up-spin state is a
weak-$*$ limit of the kink ground states
upon translation to the left, this upper bound also works for the gap above the
kink states. 
\end{proof}
\textbf{Note:} {\it The exact gap of the spin-$1/2$ XXZ chain was previously
determined rigorously, by different means, by Koma and Nachtergaele \cite{KN1}.}

\section{Example 2 : The AKLT model}
\label{Eg2AKLT}

In this section we will show how to calculate $\varepsilon(m,n)$ for the
AKLT model. The AKLT model is a spin-1, translation invariant, nearest
neighbor model with interaction
$$
h(n,n+1) = P_2(\boldsymbol{S}_n + \boldsymbol{S}_{n+1})
$$
where $P_2$ is projection onto the spin-2 states in the tensor product of
the two spins-1. This model was introduced by Affleck, Kennedy, Lieb and
Tasaki, in \cite{AKLT1}. They calculated the ground states, proved that
the ground states are exponentially clustering, and proved the existence
of a positive spectral gap \cite{AKLT2}. However, their existence proof
gave no explicit lower bound on the gap, which is is a question of physical interest
(see e.g.\ \cite{AAH}). 
In \cite{Nac1}, Nachtergaele developed
a version of the martingale method for quantum spin systems
to obtain explicit lower bounds on the gap of many 
finitely correlated spin chains.
Finitely correlated spin chains are a
class of frustration free quantum spin chains, including the
AKLT model \cite{FNW1}.
The martingale method was originally developed by Lu and Yau \cite{LY}
to obtain the exact scaling of the spectral gap and logarithmic Sobolev
inequality constant for certain stochastic spin systems.
The key point of Lu and Yau's proof for the spectral gap was the 
introduction of a filtration, such that the Gibbs measure is a 
martingale with respect to that filtration.
This is the same idea for the case of quantum spin systems, except
in place of the Gibbs measure one uses the ground state.
The martingale property guarantees that the decomposition one
obtains for the Hilbert space is actually a direct sum decomposition.
In retrospect, this same decompositon is also the key element
of the existence proof in \cite{AKLT2}, although they
did not call it a martingale decomposition.

There are several versions of the frustration free ground states of the AKLT
model.
Affleck, Kennedy, Lieb and Tasaki first wrote them down in the Ising 
basis, as a weighted sum over certain configurations which are ``alternating
sign rows''.
These are configurations of $+$'s, $-$'s and $0$'s such that between any two $+$'s there
is a $-$ and between any two $-$'s there is a $+$.
Aside from this rule, $0$'s may occur anywhere.
Physicists refer to this as diluted Ne\'el order.
On the basis of this, it was conjectured \cite{dNR} that the AKLT state has a long
range order, for an order parameter which itself is a product of arbitrarily many terms.
Partly in order to explain this, Kennedy and Tasaki \cite{KT2} came up with a 
nonlocal unitary transformation.
One of the nice features of their transformation is the fact that
it makes the ground states into product states.
This simplifies calculations, so we shall use their representation.

There is another representation of the AKLT ground state, which we 
shall not use, but we would like to mention, which is the quantum 
Markov chain (known as matrix product representation by physicists).
Fannes, Nachtergaele and Werner \cite{FNW1} used this to prove
many interesting features of the valence bond solid states, in particular that
all the correlations could be modeled on a finite-dimensional space.
They called states with this property, which include many generalizations
of the original VBS states, finitely correlated states.
More recently, physicists Kolezhuk and Mikeska \cite{KM} have
used this approach to study spin ladders, and derived new examples
of frustration free models beyond those which were known for spin chains.

The Kennedy-Tasaki unitary transformation is
$U = \prod_{j<k} \exp(i \pi S_j^z S_k^x)$.
It makes the four ground states of the AKLT model, which are otherwise
somewhat complicated states, into product states
$\Phi_k(1,N) = \bigotimes_{n=1}^N (\phi_k)_n$, where
$\phi_{1,2} = [\ket{0} \pm \sqrt{2} \ket{+}]/\sqrt{3}$
and $\phi_{3,4} = [\ket{0} \pm \sqrt{2} \ket{-}]/\sqrt{3}$.
We denote $\Phi_i(a,b) = \bigotimes_{n=a}^b (\phi_i)_n$.
These four states are not orthogonal, but we can calculate
the Gram matrix
\begin{equation*}
M(a,b) := [\ip{\Phi_i(a,b)}{\Phi_j(a,b)}]_{i,j=1}^4
  	= [\ip{\phi_i}{\phi_j}^{b-a+1}]_{i,j=1}^4\, .
\end{equation*}
Specifically,
\begin{equation}
M(1,N) = I_4 + 3^{-N} \begin{bmatrix}
  0 & (-1)^N & 1 & 1\\ (-1)^N & 0 & 1 & 1\\
  1& 1& 0& (-1)^N \\ 1& 1& (-1)^N& 0 \end{bmatrix}\, .
\end{equation}
We denote the inverse as $W(a,b) := M(a,b)^{-1}$,
and it equals
\begin{equation}
\begin{split}
W(1,N) 
  &= (1 + 2 (-1)^N 3^{-N} - 3 \cdot 3^{-2N})^{-1} \\
  &\times\,\Bigg(I_4 - 3^{-N}\begin{bmatrix} -(-1)^N 2 & (-1)^N & 1 & 1\\ 
  (-1)^N & -(-1)^N 2 & 1 & 1\\
  1& 1& -(-1)^N 2& (-1)^N \\ 1& 1& (-1)^N& -(-1)^N 2 \end{bmatrix}\Bigg)\, .
\end{split}
\end{equation}
We can write the ground state projector as
$$
G(a,b) = \sum_{i,j=1}^4 \Phi_i(a,b) W(a,b)_{ij} \Phi_j(a,b)^\dagger\, ,
$$
where $\Phi_j(a,b)^\dagger$ is the covector dual to $\Phi_j(a,b)$.
Indeed this operator is clearly Hermitian, and it has
the correct range.
It is an easy exercise for the reader to check that it behaves correctly
on the four ground states.

We will use Lemma \ref{UsefulLemma} to calculate $\varepsilon(m,n)$.
An important point is that we
do not need to use an orthogonal basis to obtain a matrix representation.
So, instead, we use the basis
$$
\Theta_{ij}(-m,n) = \Phi_i(-m,0)\otimes \Phi_j(1,n)\, .
$$
We will calculate the matrix for $G(-m,0) G(0,n) G(-m,0)$ in this basis.
Of course
$G(-m,0)$ fixes $\Theta_{ij}(-m,n)$. 
Before calculating $G(0,n) \Theta_{ij}(-m,n)$, we note that for any $a\leq c<b$,
$\Phi_i(a,b) = \Phi_i(a,c) \otimes \Phi_i(c+1,b)$.
Hence,
\begin{align*}
&G(0,n) \Theta_{ij}(-m,n) \\
  &= \sum_{kl} \Phi_k(0,n) \otimes 
  \Big( W(0,n)_{kl} \Phi_l(0,n)^\dagger
  \Phi_i(-m,-1) \otimes \Phi_i(0) \otimes \Phi_j(1,n) \Big) \\
  &=  \sum_{kl} W(0,n)_{kl} M(0,0)_{li} M(1,n)_{lj} 
  \Phi_i(-m,-1) \otimes \Phi_k(0,n)\, .
\end{align*}
Finally,
\begin{align*}
&G(-m,0) G(0,n) \Theta_{ij}(-m,n) \\
  &= \sum_{rs} \sum_{kl} \Phi_r(-m,0)\otimes 
  \Big( W(-m,0)_{rs} W(0,n)_{kl} M(0,0)_{li} M(1,n)_{lj} \\
  &\qquad \qquad \qquad \Phi_s(-m,0)^\dagger 
  \Phi_i(-m,-1) \otimes \Phi_k(0,0) \otimes \Phi_k(1,n)\Big) \\
  &= \sum_{klrs} W(0,n)_{kl} M(0,0)_{li} M(1,n)_{lj} 
  W(-m,0)_{rs} M(0,0)_{sk} M(-m,-1)_{si} \Theta_{rk}\, .
\end{align*}
Let $\mathcal{A}$ be the $16\times 16$ matrix with 
$$
\mathcal{A}(ij,rk) = \sum_{ls} W(0,n)_{kl} M(0,0)_{li} M(1,n)_{lj} W(-m,0)_{rs}
  M(0,0)_{sk} M(-m,-1)_{si}\, .
$$
The top four eigenvalues are 1, corresponding to the four-fold
degenerate ground state space $\GS(-m,n)$.
Therefore, $\varepsilon(m,n)$ is the fifth largest eigenvalue.

To diagonalize $\mathcal{A}$,
we observe that while $U$ has obscured the $\textrm{SU}(2)$ symmetry
of $\widetilde{H}_N$, it is still there. It's just no longer a ``local'' symmetry.
Each of $\GS(-m,0)$ and $\GS(1,n)$ is equal to 
$V(3) \oplus V(1)$.
Thus, the space on which $\mathcal{A}$ acts is
$(V(3) \oplus V(1))^{\otimes 2} = V(5) \oplus 3 V(3) \oplus 2V(1)$,
using Clebsch-Gordon.
Abusing notation, we refer to $3V(3)$ and $2V(1)$ as the 
spin-1 and spin-0 subspace.
$\mathcal{A}$ is $\textrm{SU}(2)$ symmetric since
each $G(a,b)$ is.
Using this and Schur's lemma, we know that $V(5)$ is an eigenspace
of $\mathcal{A}$ with eigenvalue $\lambda_5 \in \mathbb{Q}$.
We also know that the eigenspace of $\mathcal{A}$ for eigenvalue 1
is $\mathcal{G}(-m,n) \equiv V(3) \oplus V(1)$.
Let $V(1)'$ be the orthogonal complement of $\mathcal{G}(-m,n) \cap 2V(1)$
in $2V(1)$. 
Since this is one dimensional, it is an eigenspace for $\mathcal{A}$
with eigenvalue $\lambda_1 \in \mathbb{Q}$.
Let $W$ be the orthogonal complement of 
$\mathcal{G}(-m,n) \cap 3V(3)$ in $3 V(3)$.
Then $\mathcal{A}$ restricted to this block, and restricted
to an eigenspace of $S^3$, is a $2 \times 2$ matrix. 
There are two eigenvalues $\lambda_{3,\pm}$, which are irrational,
but solvable by radicals.
So $\mathcal{A}$ is solvable by hand, and the eigenvalues are
\begin{align*}
\lambda_5(m,n) &= \frac{1}{9} \cdot \frac{(1-(-1/3)^m)(1-(-1/3)^n)}{(1-(-1/3)^{m+1})(1-(-1/3)^{n+1})}\, ,\\
\lambda_1(m,n) &= \frac{1}{9} \cdot \frac{(1-(-1/3)^{m-1})(1-(-1/3)^{n-1})}{(1-(-1/3)^{m+1})(1-(-1/3)^{n+1})}\, ,\\
\lambda_{3\pm}(m,n) &= \frac{1}{9}\cdot
   \frac{1+(-1/3)^m+(-1/3)^n-19(-1/3)^{m+n+2} \pm 2 \sqrt{R(m,n)}}
  {(1-(-1/3)^{m+1})(1-(-1/3)^{n+1})}\\
R(m,n) &= [(-1/3)^m-(-1/3)^n]^2-128(-1/3)^{2m+2n+4} \\
  &\qquad
    + 4 (-1/3)^{m+n+2}(1+(-1/3)^m)(1+(-1/3)^n)\, .
\end{align*}
To calculate $\varepsilon_{1,1}$, we set $n=1$ and observe that the 
maximum eigenvalue occurs for $m=1$, $\lambda_5(1,1)=1/4$.
Since $\gamma(2)=1$, Corollary \ref{MnCor} gives
$$
\inf_{N\geq 2} \gamma(N) \geq 1 - \frac{\sqrt{3}}{2} \approx 0.133975\, .
$$
If we instead use $\varepsilon_{3,3}$, we must numerically diagonalize the
six-site chain to calculate $\gamma_6$.
We have done this using Lanczos, and the answer is
$\gamma_6 = 0.3985 (\pm 0.00005)$.
Taking $n=3$ we find $\varepsilon(3,3) = \lambda_5(3,3) = (7/20)^2$.
So we obtain
$$
\inf_{N\geq 2} \gamma(N) \geq 0.137194\, .
$$

Several authors have calculated variational upper bounds for the spectral gap above
the AKLT ground states.
One of these authors was Knabe \cite{Knabe} who not only calculated variational upper bounds,
but obtained rigorous lower bounds, too. 
F\'ath and Soly\'om \cite{FS} showed Knabe's trial states to be equivalent to
the single mode approximation states of Arovas, Auerbach and Haldane \cite{AAH},
and themselves gave a third interpretation as hidden domain walls.
These all give an upper bound of $\gamma(N) \leq 10/27 = 0.370\dots$.
Our lower bound is not spectacular, but at least has the correct order.
By yet different methods, Fannes, Nachtergaele and Werner~\cite{FNW3} obtained the excellent 
lower bound $\gamma(N) \geq 3/10$.


\section*{Acknowledgements}
It is a pleasure to thank Bruno Nachtergaele for several fruitful discussions.
S.S. is a National Science Foundation Postdoctoral Fellow. 

\bibliographystyle{hamsplain}
\bibliography{mart}

\end{document}